\documentclass[12pt]{article}
\pdfoutput=1
\usepackage{amsfonts,amsmath,amssymb,amsthm}
\usepackage{authblk} 
\usepackage{enumitem}
\usepackage[numbers]{natbib}
\usepackage[dvips]{graphicx}
\usepackage{xcolor}
\usepackage[cp1251]{inputenc}
\usepackage[T2A]{fontenc}
\usepackage[english]{babel}

\newcommand{\neigh}{\mathcal{N}}

\DeclareMathOperator{\One}{\mathbf{1}}

\begin{document}

\author[1,2,3,*]{A. Shapoval}
\author[4]{B. Shapoval}
\author[3,2]{M. Shnirman}
\affil[1]{University of Nova Gorica, Nova Gorica, Slovenia}
\affil[2]{HSE University, Moscow, Russia}
\affil[3]{Institute of Earthquake Prediction Theory RAS, Moscow Russia}
\affil[4]{University of Colorado Boulder, USA}
\affil[*]{\small corresponding author: abshapoval@gmail.com}

\title{$1/x$ Probability Distribution in a Close Proximity of the Bak--Tang-Wiesenfeld Sandpile}

\date{\today}


\maketitle

\begin{abstract}
The mechanism of self-organized criticality is based on a steady slow loading and a quick huge stress-release.  We add the clustering of the events in space and time to the Bak--Tang--Wiesenfeld cellular automaton and obtain the truncated $1/x$ probability distribution of the events over their sizes.


\end{abstract}

\section{Introduction}


Introducing the phenomenon of self-organized criticality (SOC) and proposing
its explanation with a simple cellular automaton,
Bak, Tang, and Wisenfeld (BTW) wrote ``we believe that the new concept of
self-organized criticality can be taken much further and might be
\emph{the} underlying concept for temporal and spatial scaling in a wide
class of dissipative systems with extended degrees of freedom''~\cite{btw87}.
Researchers relate SOC to the property of a system to evolve into a critical
state without tuning any parameter.
The critical state is associated with the power-laws exhibited by the system.
The absence of adjustable parameters such as the temperature or magnetization
distinguishes the SOC systems from the systems which generate
the critical dynamics at the phase transition.
The existence of SOC in the original model is provided for by a BTW mechanism characterized
by steady slow loading, the conservative transport of stress from 
the overloaded locations, and the quick stress-release at the system
boundary~\cite{rios1999}.

The expectations of BTW came true.
Examples of SOC were claimed to discover in such different real systems
and processes as earthquake formation, forest fires, armed conflicts,
the functioning of brains, and the development of cities~\cite{bak2013nature,markovic2014power}.
The underlying systems exhibit power-law probability distributions of the events
over their sizes without external adjustments.
Nevertheless, the power-law exponents
usually depend on the features of sub-systems 
(f.~.e., seismic faults, geographical regions, and stellar types in the cases of
earthquakes, forest fires, and stellar flares, respectively,
\cite{malamud1998forest, aschwanden2021self}), thus leaving
the question regarding the extent to which the underlying systems are
self-organized to be open.

Modeling of real-life systems characterized by power-laws at the critical
state can be potentially performed with modifications
of the original BTW model that involve various ways of stress propagation
including its directed transportation and quenched disorder
\cite{dhar1989exactly,karmakar2005precise}
and implement the BTW mechanism on different spaces including fractals and networks~\cite{kutnjak1996,goh2003,shapoval2012btw,zachariou2015}.
The value of the exponent $\tau$ characterizing the power-law segment
$x^{-\tau}$ of the size-frequency relationship has been obtained numerically
for various models;
rigorous proofs have been obtained for some of them~\cite{dhar2006theoretical}.

Changes in the details of the steady loading or transport mechanism
conserve the exponent $\tau\approx 1.20$, known~\cite{priezzhev1996} for
the BTW sandpile, for its deterministic isotropic modifications
\cite{biham2001e}.
A turn to stochastic transport in isotropic sandpiles switches the exponent
to $\tau \approx 1.27$~\cite{manna1991two, biham2001e, shapoval2005cross}.
the nature of self-organized criticality is captured by
the independence of the power-laws on model details
and the existence of just a few exponents within a broad class of isotropic
sandpiles on the square lattice.
This imposing feature of the isotropic sandpiles, nevertheless,
reduces the range of its direct applications to real systems
because the latter exhibit various power-law exponents.

The purpose of this paper is a BTW-mechanism extension
that allows to tune the power-law exponent
and belongs to a ``narrow neighborhood'' of the original BTW sandpile,
thus compromising between a certain refusal from self-organization
and keeping the mechanism staying behind it.
With applications in mind, we weaken two following idealizations of
the BTW mechanism:
the complete separation of the slow and quick times scales
and, as a consequence, the impossibility to combine close in space
and time events into mega-events.
Our design of the isotropic BTW mechanism on the square lattice
will lead to $\sim 1/x$ size-frequency relationship.

\section{Model and Results}

\paragraph{Definitions.}
As in the original BTW model, we define the model dynamics
on a square lattice.
The cells of the lattice are numbered from
$1$ to $A = L^2$, where $L \in \mathbb{N}$ is the lattice length.
Each non-boundary cell $i$ shares a common side with $4$ adjacent cells.
These $4$ cells form the set $\neigh_i$ of the neighbors of the cell $i$.
The boundary cells have $3$ or $2$ (in the case of the corner cell)
neighbors.

For any $i$ an integer $h_i$ is associated with the cell $i$.
This $h_i$ is interpreted as the number of grains in the cell $i$
or as the height of the pile located in $i$.
The set of the grains $\{h_i\}_{i=1}^A$ forms the configuration.
A cell $i$ is stable if its  height $h_i < H$, where
$H$ is a threshold.
The configuration $\{h_i\}_{i=1}^A$ is stable if all the cells are stable.
The dynamics are given by changes from one stable configuration
to another according to the following procedure.

\paragraph{Avalanches and their size.}
At each time moment
$N = N_L$
different cells
$i_1$, $\ldots$, $i_N$
are chosen at random.
Their heights are increased by $1$:
\begin{equation}
  h_i \longrightarrow h_i + 1,
  \quad
  \forall i \in \{i_1, \ldots, i_N\}.
  \label{e:increase}
\end{equation}
If none of them attains the threshold $H$, nothing more occurs at this
time moment.
If at least a single height attains the threshold $H$, the grain transport
starts: unstable cells pass $H$ grains equally to the neighbors.
Formally, for any $i$ with $h_i = H$,
\begin{gather}
  h_i \longrightarrow h_i - H
  \label{e:drop}
  \\
  h_{j} \longrightarrow h_{j} + 1
  \quad\forall j \in \neigh_i.
  \label{e:pass}
\end{gather}
Let us say that each unstable cell generates an avalanche.
If $n$ unstable cells $\{i_1,\ldots,i_n\}$, $n \le N$,  appear
as a result of the grain adding at the time $t$, then $n$ avalanches
$a_{i_1,t}$, $\ldots$, $a_{i_n,t}$ occur at $t$.
At the beginning, each avalanche $a_{i_k,t}$, $k=1,\ldots,n$,
``propagates'' to a single cell,
namely, the origin $i_k$ that generates the avalanche.
At this moment the size $s_{i_k,t}$ of each avalanche is set to $0$.
The unstable cells $i_1$, $\ldots$, $i_n$ and their neighbors update the heights
in line with~\eqref{e:drop}, \eqref{e:pass} simultaneously.
The size $s_{i_k,t}$ of the avalanches $a_{i_k,t}$ is increased
from $0$ to $1$.
The updates can induce instability in other cells.
New unstable cells are associated with just those avalanches
that propagate to them.
In other words, if an unstable cell $j$ obtained a grain from
a cell $j'$ associated with the avalanche $a_{i,k}$,
then $j$ is also associated with $a_{i,k}$.
If two (or more) avalanches propagate to $j$ (i.~e., pass a grain to $j$),
then the choice of the avalanche to be assigned to $j$ is performed at random.
Each update induced by the instability of the cell associated with
the avalanche $a_{i_k,t}$ results in the increase of its size
$s_{i_k, t}$ by $1$, $k = 1,\ldots, n$.
The updates ruled by~\eqref{e:drop} and~\eqref{e:pass} occur while there
are unstable cells.
As soon as $h_i < H$ for all cells $i$ (i.~e., a stable configuration is
attained), the next time moment begins.

Note that a cell can attain the threshold $H$ several times within
a single time moment.
The correspondence to the avalanche is determined when the cell
becomes unstable.
The result of the determination can differ from case to case.

\paragraph{Mega-avalanches and their size.}
We note that the above dynamics extends the original BTW model with $N = 1$
to the case of $N > 1$.
The extension results in several avalanches spreading simultaneously.
Resolving this ambiguity,
we merge the avalanches that are close in space and time into
the mega-avalanches and focus on the probability distribution of
the mega-avalanches.
A mega-avalanche consists of a single avalanche if this avalanche is not
merged with another avalanche.

In the current version of the model,
the merging rule is formulated only with
the origin and the size of the avalanches and the time of their observation
allowing for random factors
to reduce the time of the computer simulation.
The proximity between the avalanches is found through the comparison of the Manhattan distance
(the sum of the absolute differences of the Cartesian coordinates)
$\pmb{\rho}$ with an appropriate function of the avalanches'
sizes.

To formalize the rule, we introduce the characteristic (two-state) function
$\One_{\mathsf{condition}} = \One_{\mathsf{condition}}(i_1, t_1, i_2, t_2)$
that attains $1$ if the $\mathsf{condition}$ holds
and $0$ otherwise.
Let $U\sim\mathbf{Uni}(0,1)$ be a uniform $[0, 1]$ random variable.
Then the inequality
\begin{equation}
  \One_{\pmb{\rho}(i_1, i_2) <
    C'L (s_{i_1,t_1}^d + s_{i_2,t_2}^d)
  } \cdot \One_{|t_1 - t_2| \le T}
  +
  \One_{U < p} \cdot \One_{t_1 = t_2}
  > 0
  \label{e:keycondition}
\end{equation}
underlies the merging of $a_{i_1,t_1}$ and $a_{i_2,t_2}$,
where $p \in [0, 1]$, $T \ge 0$, $C' > 0$, and $d > 0$ are the parameters.
We fix $C' = 0.025$ and $d = 0.33$, taking them from a range of affordable
values. The specific choice affects the other parameters that result
in the scale-free distribution $f_L(s)$.

The specific choice of the parameters $T = 0$ and $p = 0$
simplifies~\eqref{e:keycondition} to 
\begin{equation}
  \pmb{\rho}(i_1, i_2) < C'L (s_{i_1,t_1}^d + s_{i_2,t_2}^d),
  \quad
  t_1 = t_2.
  \label{e:spaceclust}
\end{equation}
The switch to positive values of $p$ admits a random merging of the
avalanches. Positive integers $T$ allow to coalesce the avalanches
observed at subsequent time moments. As we will see, a gradual increase
in $T$ from zero is required rather than the jump to $1$.
This leads us to the fractional values of $T \in (0, 1)$ and
the probabilistic nature of the inequality $|t_1 - t_2| \le T$.
This inequality is claimed to hold with certainty if $t_1 = t_2$
and with probability $T$ if $|t_1 - t_2| = 1$.

If the avalanches $a_{i_1,t_1}$, $\ldots$, $a_{i_k,t_k}$, $k \ge 1$, form the
mega-avalanche $a$, then the size $s = \mathsf{size}(a)$ of $a$
is the sum of the corresponding
sizes:
$s = s_{i_1,t_1} + \ldots + s_{i_k,t_k}$.
The origin of the mega-avalanches is the weighted average of the origins
of the contributing avalanches, where the weights are proportional
to the sizes of the avalanches.

\paragraph{Probability distribution of the mega-avalanches.}
Let $f_L(s)$ be the empirical density of the mega-avalanches
occurred on the $L \times L$-lattice
with respect to their sizes $s$
and
$F_L(s) = \#\{a: \sigma=\mathsf{size}(a) \in [s / \Delta s, s \Delta s)\}
  / \#\{a: \sigma =\mathsf{size}(a)> 0\}$ 
be the proportion of the mega-avalanches with the size located between
$s/\Delta s$ and $s \Delta s$,
where $\Delta s = 1.2$ is chosen in the graphs.
Note that the focus on $F_L(s)$ instead of $f_L(s)$
increases the exponent of the power-law from $-\tau$ to $-\tau+1$,
as it follows from the integration:
\[
  \int_{s/\Delta s}^{s\Delta s} \pmb{\sigma^{-\tau}}\,d\sigma
  =
  \begin{cases}
    \frac{1}{1-\tau}
    \big(
      (\Delta s)^{1-\tau} - (\Delta s)^{\tau-1}
    \big) \cdot \pmb{s^{1-\tau}}, & \text{if $\tau \ne 1$,}
    \\
    2\ln\Delta s \cdot \pmb{s^0}, & \text{if $\tau = 1$.}
  \end{cases}
\]

\begin{figure}[ht]
  \hbox to \hsize{\hfil
    \includegraphics[width = 0.75\textwidth]{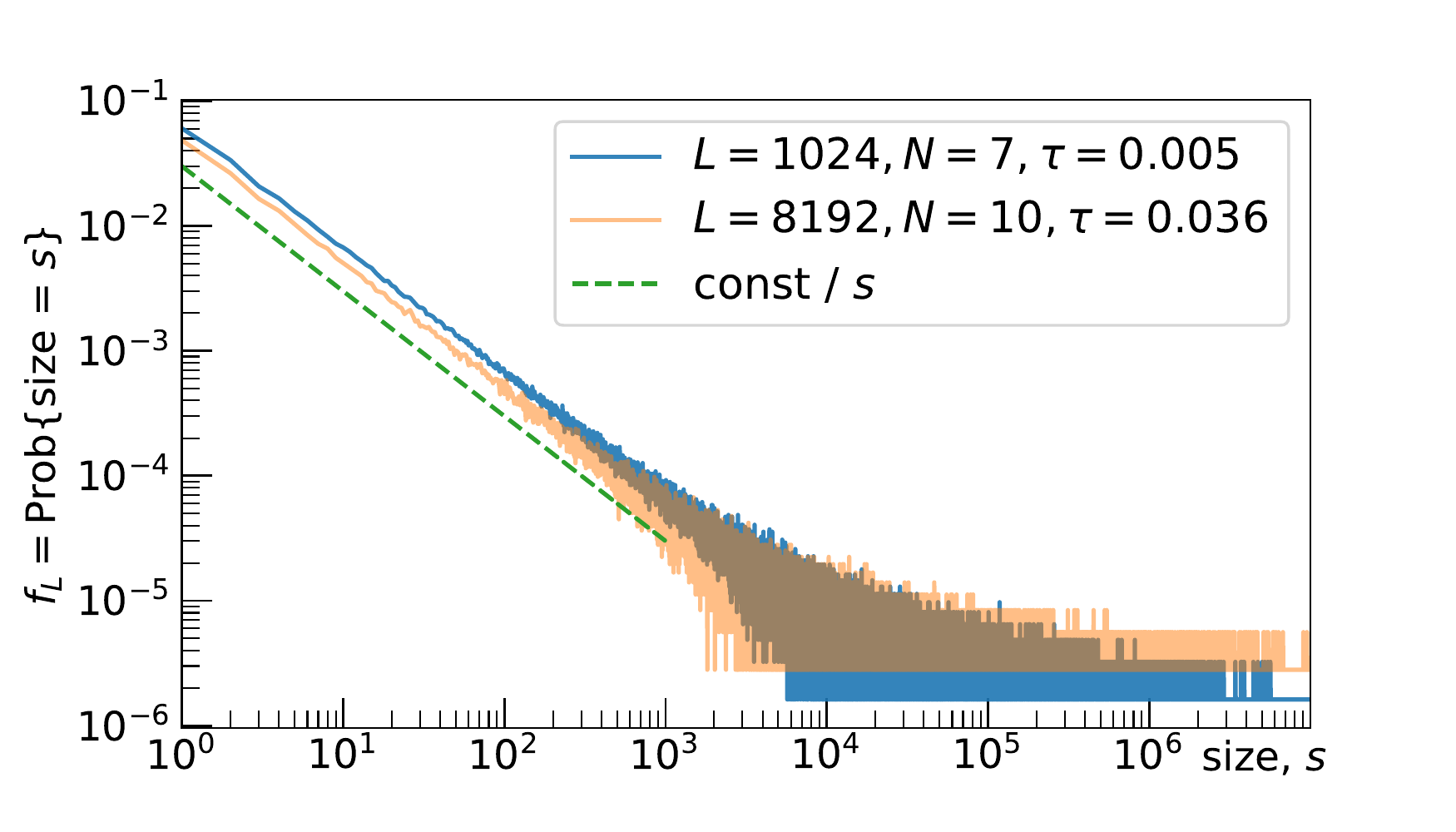}
  \hfil}  
  \caption{%
    \small
    The probability density function $f_L(s)$ of mega-avalanche's sizes;
    the part with $s > 10^7$ is omitted.
  }
  \label{f:dens}
\end{figure}

\begin{figure}[ht]
  \hbox to \hsize{\hfil
    \includegraphics[width = 0.99\textwidth]{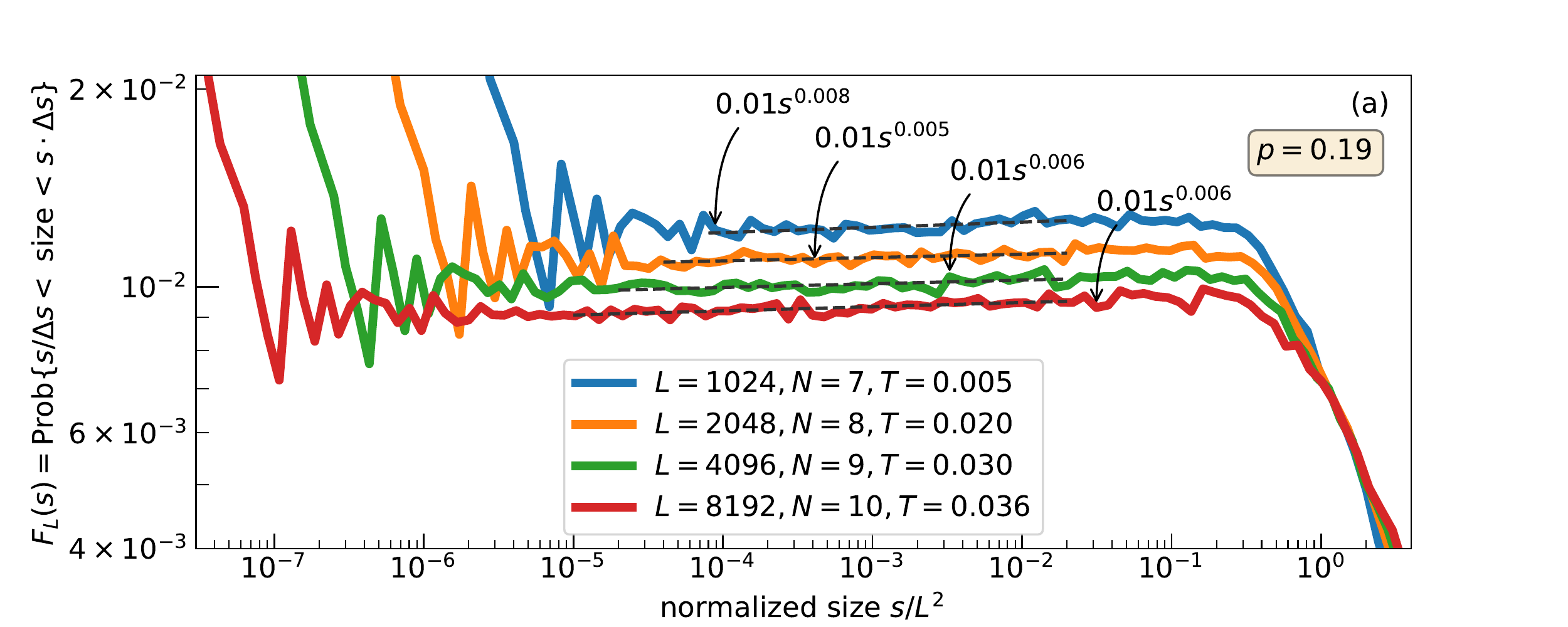}
    \hfil
  }
  \hbox to \hsize{\hfil
    \includegraphics[width = 0.99\textwidth]{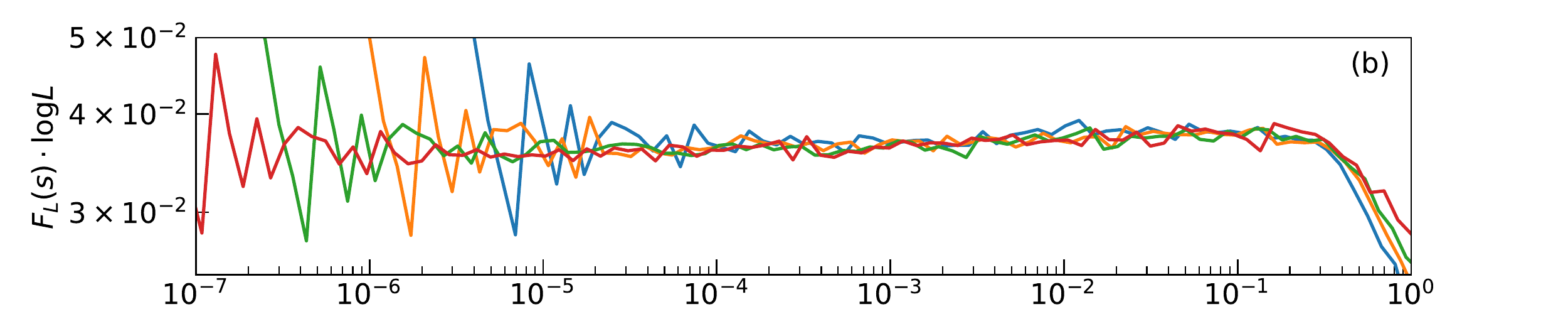}
    \hfil
  }
  \caption{%
    \small
    Power-law segment of the probability distributions of avalanches' sizes,
    where the length of the horizontal bins increases exponentially
    (which rises the power-law exponent by $1$)
    and the vertical axis is (b) and is not (a) normalized.
  }
  \label{f:area1}
\end{figure}

We have sampled the data for the empirical functions $f_L(s)$ and $F_L(s)$ for
$5 \cdot 10^5$ subsequent time moments for all lattices.
Sampling is performed after some transient period to
let the system reach the steady state and
eliminate the dependence on the initial conditions.
The graph of $f_L(s)$
is too noisy at the right to illustrate the full power-law segment
(Figure~\ref{f:dens} exhibits $f_L(s)$ found with $L=1024$ and
$L=8192$).
On the contrary, gathering the points of $f_L(s)$ within the
exponentially growing bins into $F_L(s)$, we give the relevant pattern
of the power-law segment up to the abrupt bend down in the log-log scale,
Figure~\ref{f:area1}.
The scaling $s \to s/L^{2}$ normalizes the right endpoint of the power-law
segment, Figure~\ref{f:area1}.

All four graphs of Figure~\ref{f:area1} follow an almost flat step that
is turned to a quick decay at the right.
We emphasize that the power-law exponents are found with the maximum
likelihood method applied to the empirical probability density $f_L(s)$.
These exponents $\tau = \tau_L$ are (very) close to $1$.
The best fits to $F_L(s)$ with these exponents increased by $1$ are shown
in Figure~\ref{f:area1},
where the values $1-\tau$ are written next to the curves.
These values suggest that the graphs are almost flat, which is, indeed,
the case.

The power-law segments are collapsed after the transformation of the axis:
$s \to s/L^2$, $F_L \to F_L \log L$ (Figure~\ref{f:area1}b). 
The logarithmic correction of the vertical axis is caused by the proximity
of the probability density to $1/s$-segment, the power-law scaling of
the right endpoint $s^*$ of this segment, and a fast decay of $F_L$ at the
right from $s^*$. Then the integration of the density 
$f_L(s) = C_L / s$ over $[1, +\infty]$ results in the estimate
$C_L \cdot c \log L \approx 1$, which implies $C_L \sim 1 / \log L$.
The fact that the transformation $s \longrightarrow s/L^2$ of
the horizontal axis normalizing the right endpoint of the power-law segment
does not allow to collapse the tails is inherited from the BTW sandpile
(because of its multifractal scaling \cite{tebaldi1999m}).

\begin{figure}[ht]
  \hbox to \hsize{\hss
    \includegraphics[width = 0.65\textwidth]{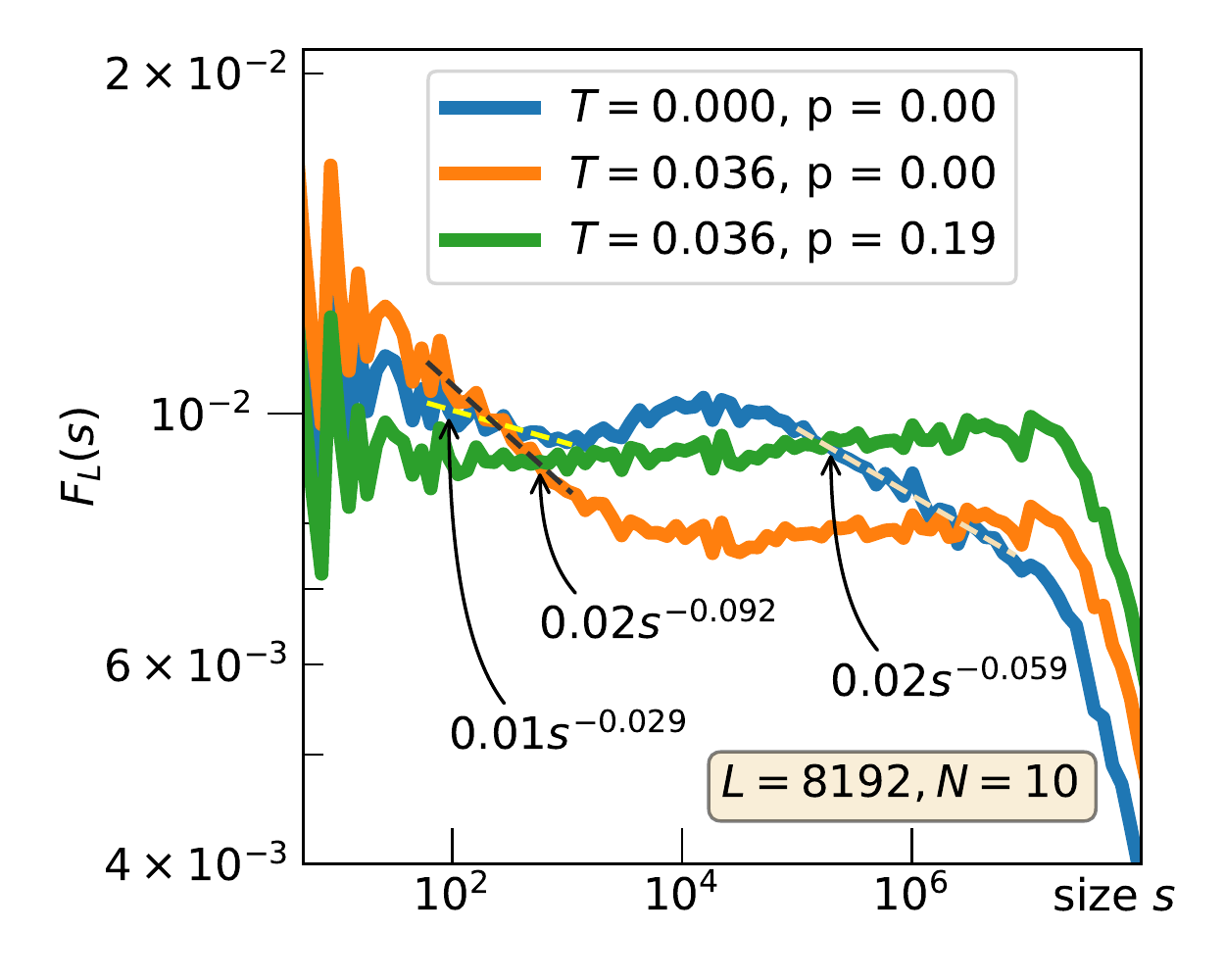}
    \hss
  }
  \caption{%
    \small
    A part of $F_L(s)$ computed with the parameters reported in the legend
    and $s^{1-\tau}$-fits.
  }
  \label{f:8192}
\end{figure}

The transition from the BTW power-law to approximately $1/s$
truncated probability distribution of the mega-avalanches
is performed with the logarithmic extra-loading $N \sim \log L$
and the probabilistic merging of the spatio-temporal clusters
of avalanches into the mega-avalanches.
The logarithmic extra-loading $N \sim \log L$ itself
with deterministic merging defined by~\eqref{e:spaceclust}
conserves the density of the grains at its critical level
(not supported by graphs) and creates two power-law parts of $F_L(s)$
The left part extends to the size of approximately $3000$ for all values
of $L$
(as seen with the blue curve in Figure~\ref{f:8192}).
Without illustration of the dependence on $L$,
we just note that
the right endpoint of second power-law part scales as $L^2$ and its
the slope becomes steeper as $L$ increases.
The introduction of the time clustering with the parameter $T > 0$
makes the right power-law part flatter in the log-log scale
(the orange curve in Figure~\ref{f:8192}).
The contraction of the gap between two consecutive values of $T$
shown in Figure~\ref{f:area1} in approximately $1.5$ times suggests
that $T$ saturates at $\approx 0.05$ as $L\to\infty$.
The right power-law part becomes narrow with the decrease of $L$,
disappearing as $L$ falls below $1024$ (not supported by graphs).
Therefore, $T$ fixed to $0.005$ for $L = 1024$ is taken as $0$
for $L = 512$ or smaller.

Interestingly, the changes in the exponent of the right power-law part
preserves the existence of the power-law at the left but alters its slope.
The return to the flat part of $F_L(s)$ is performed with the random
merging through the adjustment of the parameter $p$.
The choice of $p=0.19$ is affordable for all graphs constructed with different values of $L$.
Thus, our merging is expected to lead to $T\approx 0.05$, $p \approx 0.19$,
and $N \sim \log L$ as $L$ goes to infinity.

\section{Discussion and Conclusion}

\begin{figure}[!h]
  \hbox to \hsize{\hfil
    \includegraphics[width = 0.75\textwidth]{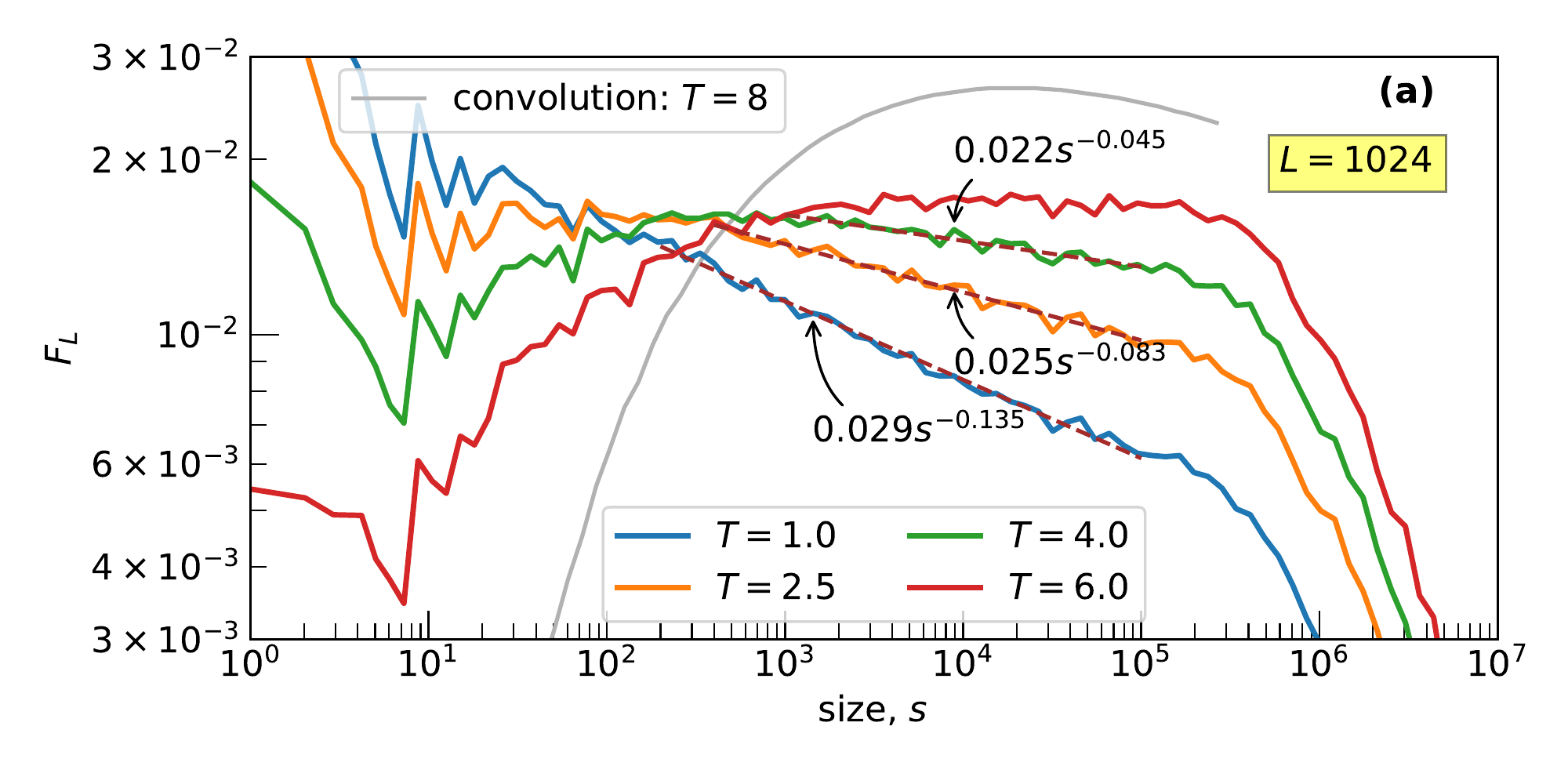}
  \hfil}
  \vspace{-4mm}
  \hbox to \hsize{\hfil
    \hspace{4mm}\includegraphics[width = 0.72\textwidth]{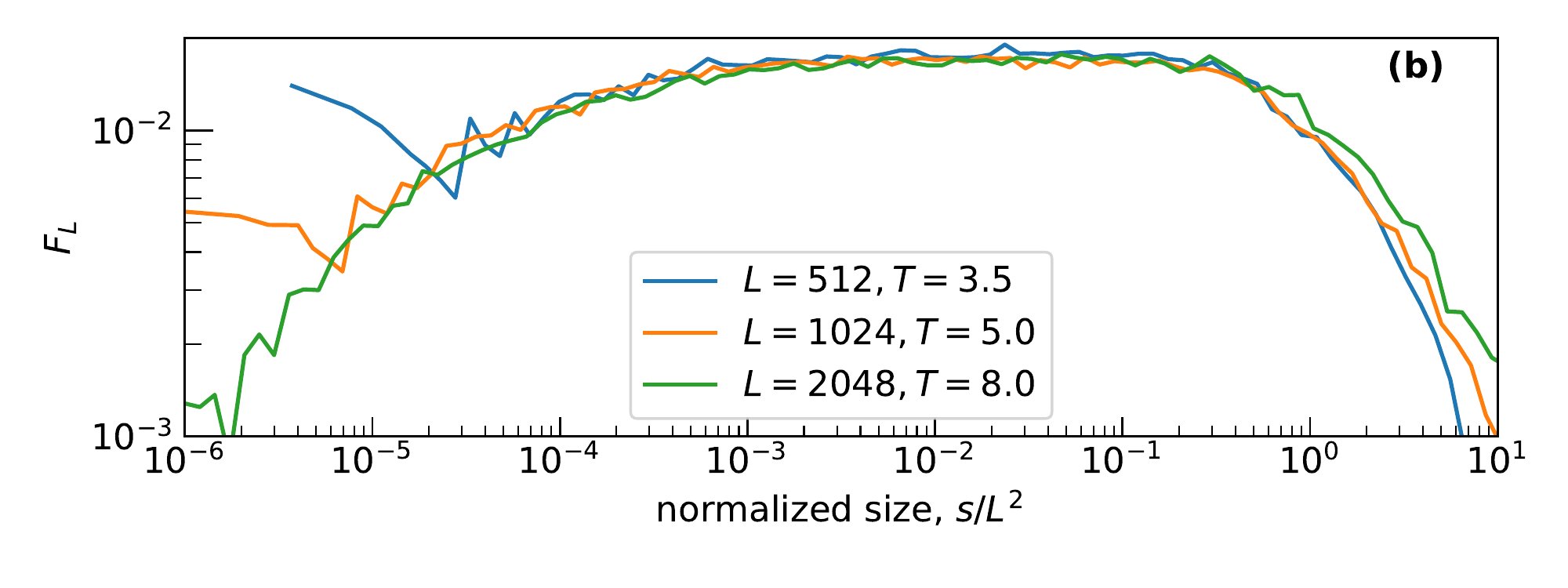}
  \hfil}
  \caption{%
    \small
    A segment of the probability distributions $F_L(s)$
    (summed up over exponentially growing bins)
    of the sizes of the mega-avalanches obtained through the merging of
    the avalanches occurred within $T$ consecutive time moments
    in the classical BTW model.
    $L^2$-normalization of the sizes collapses $F_L(s)$ with adjusted $T$, (b).
    The gray curve (in (a)) is obtained through the convolution of $8$
    probability densities $\sim s^{-1.20}$ with the support $[1, 1024]$.
  }
  \label{f:btwcombination}
\end{figure}

We insist that our approach principally differs from the two following
simple constructions: the summation of the independent power-law
random variables and merging of avalanches, which are adjacent in time,
in the original BTW model.
The first construction leads to the probability density, which is concave
in the log-log scale, tending to the power function at the right
part of the graph
(the grey curve in Figure~\ref{f:btwcombination} through $8$ convolutions,
i.~e., the summation of $8$ independent $1/x^{1.20}$ random variables
with the support on $[1, 1024]$).
The second construction can be defined through the coalescence 
of avalanches occurred during $T$ subsequent time moments.
The uncertainty with fractional values $T$ is resolved with
a probabilistic rule (say, if $T = 2.5$ and $a_{t}$ is not merged with
  $a_{t-1}$, then the avalanches $a_{t}$, $a_{t+1}$, and $a_{t+2}$ are
  combined with certainty, whereas the avalanche $a_{t+3}$
is added with the probability of $0.5$).
This modification of the BTW model
preserves the power-law segment that \emph{does not}
extend to the right with the growth of the system.
The power-law part of $F_L(s)$ constructed for the different values of $L$
is collapsed after the normalization of the size by the lattice area,
Figure~\ref{f:btwcombination}.

The paper gives evidence
that the $1/x$ power-law is feasible with isotropic extensions of
the BTW sandpile (Figure~\ref{f:area1}).
The extension is constructed with the stress accumulation,
proportional to $\log L$,
and the coalescence of the avalanches propagated closely in space and time.
Such a coalescence is known, for example, in seismology, as
the stress accumulation and the earthquakes themselves occurred in the slow
and quick time respectively are not completely separated~\cite{kanamori2003}.

The details of the construction are likely to be designed in various ways.
Proposed minor deviations from the BTW model through the parameter domain
preserve the critical density of the grains and the power-law size-frequency
relationship for the
mega-avalanches over the majority of feasible sizes (Figure~\ref{f:8192}).
The adjustment of the parameters pulls the exponent $\tau$ towards $1$
(through a weak time clustering, parameter $T$)
and corrects the slope of the restricted left part to fit
the whole power-law segment (with the random coalescence in space, parameter $p$).
Thus, our approach does not require any tuning of the dissipation-to-loading
ratio as in attempts to relate self-organized criticality to 
the phase transition modeling~\cite{dickman1998s}
but controls the universality class of the sandpile and
might lead to adjustable power-law exponents in a neighborhood of $1$.
This would improve our
understating of real-life self-organized critical phenomena.

{\small \linespread{1.0} \bibliographystyle{abbrvnat}     \bibliography{sandpile} 

\begin{thebibliography}{20}
\providecommand{\natexlab}[1]{#1}
\providecommand{\url}[1]{\texttt{#1}}
\expandafter\ifx\csname urlstyle\endcsname\relax
  \providecommand{\doi}[1]{doi: #1}\else
  \providecommand{\doi}{doi: \begingroup \urlstyle{rm}\Url}\fi

\bibitem[Aschwanden and G{\"u}del(2021)]{aschwanden2021self}
M.~J. Aschwanden and M.~G{\"u}del.
\newblock Self-organized criticality in stellar flares.
\newblock \emph{The Astrophysical Journal}, 910\penalty0 (1):\penalty0 41,
  2021.

\bibitem[Bak(2013)]{bak2013nature}
P.~Bak.
\newblock \emph{How nature works: the science of self-organized criticality}.
\newblock Springer Science \& Business Media, 2013.

\bibitem[{Bak} et~al.(1987){Bak}, {Tang}, and {Wiesenfeld}]{btw87}
P.~{Bak}, C.~{Tang}, and K.~{Wiesenfeld}.
\newblock {Self-organized criticality: an explanation of 1/f noise}.
\newblock \emph{{Phys. Rev. Lett.}}, 59:\penalty0 381--383, 1987.
\newblock \doi{10.1103/PhysRevLett.59.381}.

\bibitem[Biham et~al.(2001)Biham, Milshtein, and Malcai]{biham2001e}
O.~Biham, E.~Milshtein, and O.~Malcai.
\newblock Evidence for universality within the classes of deterministic and
  stochastic sandpile models.
\newblock \emph{Phys. Rev. E}, 63\penalty0 (6):\penalty0 061309, 2001.

\bibitem[De~Los~Rios and Zhang(1999)]{rios1999}
P.~De~Los~Rios and Y.~Zhang.
\newblock Universal 1/f noise from dissipative self-organized criticality
  models.
\newblock \emph{Phys. Rev. Lett.}, 82\penalty0 (3):\penalty0 472, 1999.

\bibitem[Dhar(2006)]{dhar2006theoretical}
D.~Dhar.
\newblock Theoretical studies of self-organized criticality.
\newblock \emph{Physica A: Statistical Mechanics and its Applications},
  369\penalty0 (1):\penalty0 29--70, 2006.

\bibitem[Dhar and Ramaswamy(1989)]{dhar1989exactly}
D.~Dhar and R.~Ramaswamy.
\newblock Exactly solved model of self-organized critical phenomena.
\newblock \emph{Phys. Rev. Lett.}, 63\penalty0 (16):\penalty0 1659, 1989.

\bibitem[Dickman et~al.(1998)Dickman, Vespignani, and Zapperi]{dickman1998s}
R.~Dickman, A.~Vespignani, and S.~Zapperi.
\newblock Self-organized criticality as an absorbing-state phase transition.
\newblock \emph{Phys. Rev. E}, 57\penalty0 (5):\penalty0 5095, 1998.

\bibitem[Goh et~al.(2003)Goh, Lee, Kahng, and Kim]{goh2003}
K.-I. Goh, D.-S. Lee, B.~Kahng, and D.~Kim.
\newblock Sandpile on scale-free networks.
\newblock \emph{Phys. Rev. Lett.}, 91\penalty0 (14):\penalty0 148701, 2003.

\bibitem[Kanamori(2003)]{kanamori2003}
H.~Kanamori.
\newblock Earthquake prediction: An overview.
\newblock In \emph{International Handbook of Earthquake and Engineering
  Seismology. International geophysics series}, volume 81B, pages 1205--1216.
  Academic Press, 2003.

\bibitem[Karmakar et~al.(2005)Karmakar, Manna, and Stella]{karmakar2005precise}
R.~Karmakar, S.~Manna, and A.~Stella.
\newblock Precise toppling balance, quenched disorder, and universality for
  sandpiles.
\newblock \emph{Phys. Rev. Lett.}, 94\penalty0 (8):\penalty0 088002, 2005.

\bibitem[Kutnjak-Urbanc et~al.(1996)Kutnjak-Urbanc, Zapperi,
  Milo{\v{s}}evi{\'c}, and Stanley]{kutnjak1996}
B.~Kutnjak-Urbanc, S.~Zapperi, S.~Milo{\v{s}}evi{\'c}, and H.~Stanley.
\newblock Sandpile model on the sierpinski gasket fractal.
\newblock \emph{Phys. Rev. E}, 54\penalty0 (1):\penalty0 272, 1996.

\bibitem[Malamud et~al.(1998)Malamud, Morein, and Turcotte]{malamud1998forest}
B.~Malamud, G.~Morein, and D.~Turcotte.
\newblock Forest fires: an example of self-organized critical behavior.
\newblock \emph{Science}, 281\penalty0 (5384):\penalty0 1840--1842, 1998.

\bibitem[Manna(1991)]{manna1991two}
S.~Manna.
\newblock Two-state model of self-organized criticality.
\newblock \emph{Journal of Physics A: Mathematical and General}, 24\penalty0
  (7):\penalty0 L363, 1991.

\bibitem[Markovi{\'c} and Gros(2014)]{markovic2014power}
D.~Markovi{\'c} and C.~Gros.
\newblock Power laws and self-organized criticality in theory and nature.
\newblock \emph{Physics Reports}, 536\penalty0 (2):\penalty0 41--74, 2014.

\bibitem[Priezzhev et~al.(1996)Priezzhev, Ktitarev, and
  Ivashkevich]{priezzhev1996}
V.~Priezzhev, D.~Ktitarev, and E.~Ivashkevich.
\newblock Formation of avalanches and critical exponents in an abelian sandpile
  model.
\newblock \emph{Phys. Rev. Lett.}, 76\penalty0 (12):\penalty0 2093, 1996.

\bibitem[Shapoval and Shnirman(2005)]{shapoval2005cross}
A.~Shapoval and M.~Shnirman.
\newblock Crossover phenomenon and universality: From random walk to
  deterministic sand-piles through random sand-piles.
\newblock \emph{International Journal of Modern Physics C}, 16\penalty0
  (12):\penalty0 1893--1907, 2005.

\bibitem[Shapoval and Shnirman(2012)]{shapoval2012btw}
A.~Shapoval and M.~Shnirman.
\newblock The {BTW} mechanism on a self-similar image of a square: A path to
  unexpected exponents.
\newblock \emph{Physica A: Statistical Mechanics and its Applications},
  391\penalty0 (1-2):\penalty0 15--20, 2012.

\bibitem[Tebaldi et~al.(1999)Tebaldi, De~Menech, and Stella]{tebaldi1999m}
C.~Tebaldi, M.~De~Menech, and A.~L. Stella.
\newblock Multifractal scaling in the bak-tang-wiesenfeld sandpile and edge
  events.
\newblock \emph{Phys. Rev. Lett.}, 83\penalty0 (19):\penalty0 3952, 1999.

\bibitem[Zachariou et~al.(2015)Zachariou, Expert, Takayasu, and
  Christensen]{zachariou2015}
N.~Zachariou, P.~Expert, M.~Takayasu, and K.~Christensen.
\newblock Generalised sandpile dynamics on artificial and real-world directed
  networks.
\newblock \emph{PloS One}, 10\penalty0 (11):\penalty0 e0142685, 2015.

\end{thebibliography}
}

\end{document}